\documentclass[prl,showpacs,twocolumn,amssymb]{revtex4}
\begin{document}

\title{Magnetic  Color Flavor Locking Phase in High Density QCD}
\author{Efrain J. Ferrer}
\author{Vivian de la Incera}
\affiliation{Department of Physics, Western Illinois University,
Macomb, IL 61455, USA}
\author{Cristina Manuel}
\affiliation{Instituto de F\'{\i}sica Corpuscular, CSIC-U. de
Val\`encia, 46071 Val\`encia, Spain}

\begin{abstract}

We investigate the effects of an external magnetic field in the gap
structure of a color superconductor with three massless quark
flavors. Using an effective theory with four-fermion interactions,
inspired by one-gluon exchange, we show that the long-range
component $\widetilde{B}$ of the external magnetic field that
penetrates the color-flavor locked (CFL) phase modifies its gap
structure, producing a new phase of lower symmetry. A main outcome
of our study is that the $\widetilde{B}$ field tends to strengthen
the gaps formed by $\widetilde{Q}$-charged and
$\widetilde{Q}$-neutral quarks that coupled among themselves through
tree-level vertices. These gaps are enhanced by the field-dependent
density of states of the $\widetilde{Q}$-charged quarks on the Fermi
surface. Our considerations are relevant for the study of highly
magnetized compact stars.

\end{abstract}
\maketitle
After the suggestion that three-flavor quark matter may
actually be the ground state of strong interactions
\cite{Witten:1984rs}, quark stars were postulated as possible
astrophysical objects. It is also very likely that quark matter
occupies the inner regions of neutron stars.  Our present
knowledge of QCD at high baryonic density indicates that this new
state of matter might be in a color superconducting phase (for
reviews, see \cite{reviews}).
 On the
other hand,  it is well-known \cite{Lai:2000at} that strong
magnetic fields, as large as $B \sim 10^{12} - 10^{14}$ G,  exist
in the surface of neutron stars, while in magnetars they are in
the range $B \sim 10^{14} - 10^{15}$ G, and perhaps as high as
$10^{16}$ G \cite{magnetars}.
  The
physical upper limit to the total neutron star magnetic field, as
arising from comparing the magnetic and  gravitational energies,
is of order $B \sim 10^{18}$ G \cite{Lai:2000at}. If quark stars
are self-bound rather than gravitational-bound objects this upper
limit could go higher. In this Letter we investigate the effect of
a strong magnetic field in color superconductivity, with the aim
of further studying its possible astrophysical implications.

We will start by considering three massless quarks. In this
case, it is well established that the ground state  of
high-dense QCD corresponds to the CFL (color-flavor locked) phase
\cite{alf-raj-wil-99/537}.  In this phase, quarks form spin-zero
Cooper pairs in the color- antitriplet, flavor-antitriplet
representation, thereby breaking the original $SU(3)_{{\rm
color}}\times SU(3)_{L}\times SU(3)_{R} \times U(1)_B$ symmetry to
the diagonal subgroup $SU(3)_{{\rm color}+L+R}$. One can now ask
how this scenario will change when a magnetic field is switched
on. Would the external field affect the pairing phenomena?
 In a conventional electromagnetic superconductor, since
Cooper pairs are electrically charged, the electromagnetic gauge
invariance is spontaneously broken and the photon acquires a mass
that can screen a weak magnetic field: this is the Meissner
effect. In spin-zero color superconductivity, although the color
condensate has non-zero electric charge, there is a linear
combination of the photon and a gluon that remains massless
\cite{alf-raj-wil-99/537}. The unbroken $U(1)$ group is generated,
in flavor-color space, by $\widetilde {Q} = Q \times 1 - 1 \times
Q$, where $Q$ is the electromagnetic charge generator
\cite{footnote}. Thus a spin-zero color superconductor may be
penetrated by a long-range remnant ``rotated magnetic" field
$\widetilde{B}$. In the 9-dimensional flavor-color representation
that we will use in this paper, the $\widetilde{Q}$ charges of the
different quarks are
\begin{equation}
\begin{tabular}{|c|c|c|c|c|c|c|c|c|}
  \hline
  $s_{1}$ & $s_{2}$ & $s_{3}$ & $d_{1}$ & $d_{2}$ & $d_{3}$ & $u_{1}$ & $u_{2}$ & $u_{3}$ \\
  \hline
  0 & 0 & - & 0 & 0 & - & + & + & 0 \\
  \hline
\end{tabular} \ ,
\end{equation}
in units of the $\widetilde {Q}$-charge of the electron
$\widetilde{e} = e \cos{\theta}$, where $\theta$ is the mixing
angle \cite{Litim:2001mv} (we set $\hbar=c=1$ henceforth).

Although the interaction of an external magnetic field with dense
quark matter has been investigated by several authors
\cite{magnetic-orig,iida}, they disregarded the effect of the
penetrating $\widetilde{B}$ field on the gap structure. However,
the $\widetilde{B}$ field can change the gap structure and lead to
a new superconducting phase. To understand how this can occur
notice that due to the coupling of the charged quarks with the
external $\widetilde{B}$ field, the color-flavor space is
augmented by the $\widetilde{Q}$ charge operator, and consequently
the order parameter of the CFL splits in new independent pieces.

Based on the above considerations, and imposing that the
condensates should retain the highest degree of symmetry,
 we propose the following ansatz
for the color-flavor structure of the order parameter of three
massless quarks in the presence of a magnetic $\widetilde{B}$
field:
\begin{equation}
\label{mcfl-gap} \Delta^{+}=k_{1}^+ (U_{0}-N) \Omega
_{0}+k_{2}^{+} U \Omega _{0}+k_{n}^{+} N \Omega _{0}+k_{c}^{+} U
[\Omega _{+}+\Omega _{-}] \label{1}
\end{equation}
with color-flavor matrices defined as: $U_{0}=\delta_{a}^{i}
\delta_{b}^{j}$, $U=\delta_{a}^{j} \delta_{b}^{i}$,
$N={(\delta_{a}^{1}\delta_{i}^{1}
+\delta_{a}^{2}\delta_{i}^{2})\delta_{b}^{3}\delta_{j}^{3}+(a\leftrightarrow
b,i\leftrightarrow j})$; with $a,b$ and $i,j$ denoting color and
flavor indexes respectively. The matrices $\Omega
_{0}=diag(1,1,0,1,1,0,0,0,1)$,
$\Omega_{+}=diag(0,0,0,0,0,0,1,1,0)$, and
$\Omega_{-}=diag(0,0,1,0,0,1,0,0,0)$ are $\widetilde{Q}$-charge
projectors with algebra $\Omega_{\eta }\Omega _{\eta ^{\prime
}}=\delta _{\eta \eta^{\prime }}\Omega _{\eta }$, for $\eta ,\eta
^{\prime }=0,+,-$, and $\Omega _{0}+\Omega _{+}+\Omega _{-}=1$.

An applied magnetic field reduces the flavor symmetries of QCD, as
only the $d$ and $s$ quarks have equal electromagnetic charge. Thus,
the order parameter (\ref{mcfl-gap}) implies the following symmetry
breaking pattern: $SU(3)_{\rm color} \times SU(2)_L \times SU(2)_R
\times U(1)_B \times U^{(-)}(1)_A \times U(1)_{\rm e.m.} \rightarrow
SU(2)_{{\rm color}+L+R} \times {\widetilde {U}(1)}_{\rm e.m.}$. The
$U^{(-)}(1)_A$ symmetry is connected with the current which is an
anomaly free linear combination of $s,d$ and $u$ axial currents
\cite{miransky-shovkovy-02}.  The locked $SU(2)$ corresponds to the
maximal unbroken symmetry, and as such it maximizes the condensation
energy. Notice that it commutes with ${\widetilde {U}(1)}_{\rm
e.m.}$.

 Therefore, there are 13
broken generators, 8 of which become the longitudinal components
of massive gauge bosons, and 5 remain as Goldstone bosons. One is
associated to the spontaneous breaking of baryon symmetry, one
with the breaking of the anomaly free  $U^{(-)}(1)_A$,
 and the
remaining 3 to the breaking of the chiral $SU(2)$ group. This
symmetry breaking pattern suggests that the new phase has
quantitative and qualitative differences with respect to the CFL
phase. We will call it magnetic CFL (MCFL) phase. In particular,
the MCFL phase will possess a distinctive low energy physics.

To trace back the physical origin of the new structures in (\ref{mcfl-gap})
 we should
take into account that despite the $\widetilde{Q}$-neutrality of all
the condensates, they can be composed either by neutral or by
charged quarks. Condensates formed by $\widetilde{Q}$-charged quarks
feel the field directly through the minimal coupling of the
background field $\widetilde{B}$ with the quarks in the pair. A
subset of the condensates formed by $\widetilde{Q}$-neutral quarks,
can feel the presence of the field through the tree-level vertices
that couple them to charged quarks. The gaps $\Delta^B_{A/S} \equiv
(k_{n} \mp k_{c})/{2}$ are antisymmetric/symmetric combinations of
condensates composed by charged quarks and condensates formed by
this kind of neutral quarks. The gaps $\Delta_{A/S} \equiv (k_{1}
\mp k_{2})/{2}$, on the other hand, are antisymmetric/symmetric
combinations of condensates formed by neutral quarks that do not
belong to the above subset. The only way the field can affect
$\Delta_{A/S}$ is through the system of highly non-linear coupled
gap equations. The CFL gap matrix is obtained when
$\Delta^{B}_{A/S}=\Delta_{A/S}$. In principle, the symmetries of the
problem allow for two extra independent symmetric gaps. But these
are only due to subleading color symmetric interactions, and are
formed by neutral quarks that are not coupled to charged quarks, so
they belong to the same class as $\Delta_{S}$. Thus, in a first
approach to the problem, we will consider that those can as well be
described by $\Delta_S$.

To study the MCFL phase we use a Nambu-Jona-Lasinio (NJL)
four-fermion interaction abstracted from one-gluon exchange
\cite{alf-raj-wil-99/537}. This simplified treatment, although
disregards the effect of the $\widetilde {B}$-field on the gluon
dynamics and assumes the same NJL couplings for the system with
and without magnetic field, keeps the main attributes of the
theory, providing the correct qualitative physics. We will
postpone the study within QCD for the future.

The NJL model is defined by two parameters, a coupling constant $g$
and an ultraviolet cutoff $\Lambda$. The cutoff should be higher
than the typical energy scales in the system, that is, the chemical
potential $\mu$ and the magnetic energy
$\sqrt{\widetilde{e}\widetilde{B}}$.

 The gap equation of the NJL model in coordinate space
reads
\begin{equation}
\label{gap-eq} \Delta^+ = i\frac{g^2}{4 \Lambda^2} \lambda_A^T\,
\gamma^\mu\, S_{21}(x,y) \gamma_\mu\, \lambda_A \,
\delta^{(4)}(x-y) \ ,
\end{equation}
where $\lambda_A$ and $\gamma_\mu$ are the Gell-Mann and Dirac
matrices, respectively. For simplicity, we have omitted explicit
flavor, color and spinor indexes in the equation. $S_{21}$ is the
21-component of the quark propagators in the Nambu-Gorkov
representation.

The computation of the field-dependent quark propagators is
laborious (details will be given elsewhere \cite{future}), but it
can be managed with the use of the Ritus' method, originally
developed for charged fermions \cite{Ritus:1978cj} and recently
extended to charged vector fields \cite{efi-ext}. In Ritus'
approach the diagonalization in momentum space of charged fermion
Green's functions in the presence of a background magnetic field
is carried out using the eigenfunction matrices $E_p(x)$. These
are the wave functions of the asymptotic states of charged
fermions in a uniform magnetic field and play the role in the
magnetized medium of the usual plane-wave (Fourier) functions
$e^{i px}$ at zero field. With the help of the $E_p(x)$ functions,
we first compute the propagators in momentum space, and then
transform to coordinate space adequately. Leaving aside the
color-flavor structure, the neutral quark propagators are of the
same type as for the CFL phase. The charged (positive/negative)
quark propagators in the background of a $\widetilde{B}$ field
that lies in the $\hat{z}$-axis are
\begin{equation}
\label{propag}
 \nonumber S_{21}^{(\pm)} = \frac{\Lambda^+_{(\pm)}
\gamma_5\, k_c}{ p^2_0 -(|{\bf \overline{p}}^{(\pm)}| -\mu)^2
-k^2_c} + \frac{\Lambda^-_{(\pm)} \gamma_5\, k_c}{ p^2_0 -(|{\bf
\overline{p}}^{(\pm)}| +\mu)^2 -k^2_c}
\end{equation}
where ${\bf \overline{p}}^{(\pm)} = (0, \pm \sqrt{2
|\widetilde{e}\widetilde{B}| l},p_3)$ are the spatial components
of the momentum for (positive/negative) charged quarks and the
integer number $l$ labels the Landau levels.
 In
 (\ref{propag}),
 $\Lambda^{(\pm)}_{\pm}=
(1\pm\gamma_{0}{\mathbf{\gamma}} \cdot
\widehat{\overline{\mathbf{p}}}^{(\pm)})/2$ are the energy
projectors in the ultrarelativistic limit for (positive/negative)
charged quarks in the external field, with
$\widehat{\overline{\mathbf{p}}}^{(\pm)}$ representing the
normalized charged-quark three-momentum. On the Fermi surface, the
highest occupied Landau level is obtained as $l_{\rm max} =
\Big[\frac{\mu^2}{2 |\widetilde{e} \widetilde{B}|}\Big]$, where
the bracket denotes integer part.

Since color superconductivity, although of type I, allows the
penetration of a rotated magnetic field, it is natural to expect
that the condensates made up of ${\widetilde Q}$-charged quarks
will be strengthened by the nonzero $\widetilde{B}$, because these
paired quarks have opposite ${\widetilde Q}$ charges and opposite
spins, hence parallel (instead of antiparallel) magnetic moments.
The situation here has some resemblance to the magnetic catalysis
(MC) of chiral symmetry breaking \cite{MC}, in the sense that the
magnetic field strengthens the pair formation. Despite this
similarity, the way the field influences the pairing mechanism in
the two cases is quite different. The particles participating in
the chiral condensate are near the surface of the Dirac sea. The
effect of a magnetic field there is to effectively reduce the
dimension of the particles at the lowest Landau level (LLL), which
in turn strengthens their effective coupling, catalyzing the
chiral condensate. Color superconductivity, on the other hand,
involves quarks near the Fermi surface, with a pairing dynamics
that is already $(1+1)$-dimensional. Therefore, the ${\widetilde
B}$ field does not yield further dimensional reduction of the
pairing dynamics near the Fermi surface and hence the LLL does not
have a special significance here. Nevertheless, the field
increases the density of states of the ${\widetilde Q}$-charged
quarks, and it is through this effect that the pairing of the
charged particles is reinforced by the penetrating magnetic field.
Below we will analytically show that this is indeed the case.

To solve the gap equation (\ref{gap-eq}) for the whole range of
magnetic-field strengths we need to use numerical methods. We have
found, however, a situation where an analytical solution is
possible. This corresponds to the case $\widetilde{e}\widetilde{B}
\gtrsim \mu^2/2$. Taking into account that the leading
contribution to the gap solution comes from quark energies near
the Fermi level, it follows that for fields in this range only the
LLL ($l=0$) contributes.

Using the approximation $\Delta^B_A \gg \Delta^B_S, \Delta_A$,
$\Delta_A \gg \Delta_S$, the gap equations decouple and the
equation for $\Delta^B_A$ is

\begin{eqnarray}
\label{maingeq}
\Delta^B_A & \approx & \frac{g^2}{3 \Lambda^2} \int_{\Lambda}
\frac{d^3 q}{(2 \pi)^3}
\frac{ \Delta^B_A}{\sqrt{(q-\mu)^2 + 2 (\Delta^B_A)^2 }}
\nonumber
 \\
& + & \frac{g^2 \widetilde{e}\widetilde{B}}{3 \Lambda^2} \int_{-\Lambda}^{\Lambda}
\frac{d q}{(2 \pi)^2} \frac{ \Delta^B_A}{\sqrt{(q-\mu)^2 +
(\Delta^B_A)^2 }}   \ ,
\end{eqnarray}
where the first/second term in the r.h.s. of Eq.~(\ref{maingeq})
corresponds to the contribution of $\widetilde{Q}$-neutral/charged
quark propagators, respectively. For the last one, we dropped all
Landau levels but the lowest, as we are interested in the leading
term.

The solution of Eq.~(\ref{maingeq}) reads
\begin{equation}
\label{gapBA}
\Delta^B_A \sim 2 \sqrt{\delta \mu} \, \exp{\Big( - \frac{3 \Lambda^2
\pi^2} {g^2 \left(\mu^2 + \widetilde{e} \widetilde{B} \right)}
\Big) } \ ,
\end{equation}
with $\delta \equiv \Lambda - \mu$,
to be compared with the antisymmetric CFL gap \cite{reviews}
\begin{equation}
\label{gapCFL}
\Delta^{\rm CFL}_A \sim 2 \sqrt{\delta \mu} \, \exp{\Big( -\frac{3
\Lambda^2 \pi^2} {2 g^2 \mu^2} \Big) } \ .
\end{equation}

In this approximation the remaining gap equations read
\begin{eqnarray}
\label{symgeq} \Delta^B_S & \approx & -\frac{g^2}{6 \Lambda^2} \int_{\Lambda}
\frac{d^3 q}{(2 \pi)^3}
\frac{ \Delta^B_A}{\sqrt{(q-\mu)^2 + 2 (\Delta^B_A)^2 }} \nonumber\\
& + & \frac{g^2 \widetilde{e}\widetilde{B}}{6 \Lambda^2} \int_{-\Lambda}^{\Lambda}
\frac{d q}{(2 \pi)^2} \frac{ \Delta^B_A}{\sqrt{(q-\mu)^2 +
(\Delta^B_A)^2 }}   \ ,
\end{eqnarray}

\begin{eqnarray}
\label{antisymme-gapeqapp}
 \Delta_A & \approx & \frac {g^2}{4 \Lambda^2}  \int_{\Lambda} \frac{d^3q}{(2 \pi)^3}
 \Big( \frac {17}{9} \frac{\Delta_A}{ \sqrt{ (q - \mu)^2 +\Delta_A^2 }}
\nonumber \\
& + &  \frac{7}{9} \frac{\Delta_A}{ \sqrt{(q-\mu)^2 + 2 (\Delta^B_A)^2 }  }
 \Big) \ ,
 \end{eqnarray}
and

\begin{eqnarray}
\label{symme-gapeqapp}
 \Delta_S & \approx &  \frac {g^2}{18 \Lambda^2}  \int_{\Lambda} \frac{d^3q}{(2 \pi)^3}
\Big(  \frac{\Delta_A}{ \sqrt{ (q - \mu)^2 +\Delta_A^2 }} \nonumber
\\
& - &   \frac{\Delta_A}{ \sqrt{(q-\mu)^2 + 2 (\Delta^B_A)^2 } }
\Big) \ .
 \end{eqnarray}

We express below the solution of these gap equations as
ratios over the CFL antisymmetric and symmetric gaps
\begin{equation}
 \frac{\Delta_A}{\Delta^{\rm CFL}_A}  \sim \frac{1}{2^{(7/34)}} \exp{\Big(-\frac{36}{17 x}  + \frac{21}{17}\frac{1}{x (1 + y)} + \frac{3}{2x}
\Big) } \ ,
\end{equation}
where $x \equiv g^2 \mu^2/\Lambda^2 \pi^2$, and $y \equiv
\widetilde{e}\widetilde{B}/\mu^2$, and
\begin{eqnarray}
\frac{\Delta^B_S}{\Delta^{\rm CFL}_S} & \sim &
\frac{\Delta^B_A}{\Delta^{\rm CFL}_A} \left(
\frac{3}{4} + \frac{9}{2 x \ln{2}} \frac{y -1}{y+1} \right) \ ,\\
\frac{\Delta_S}{\Delta^{\rm CFL}_S} & \sim & \frac{\Delta_A}{\Delta^{\rm
CFL}_A} \frac 32 \left(1 - \frac{4}{1+y} \right) \ .
\end{eqnarray}

Note that our analytic solutions are only valid at strong magnetic
fields. The lower value $\widetilde{e}\widetilde{B} \sim \mu^2/2$
corresponds to $\widetilde{e}\widetilde{B} \sim (0.8 - 1.1) \cdot
10^{18}$G, for $\mu \sim 350- 400$ MeV. For fields of this order and
larger the $\Delta^B_A$ gap is larger than $\Delta^{\rm CFL}_A$ at
the same density values. Note also that for
$\widetilde{e}\widetilde{B}\gtrsim \mu^2$, $\Delta_{A/S}^B$ grow and
$\Delta_{A/S}$ tend to lower, and they clearly split. How fast or
slowly they do depends very much on the values of the NJL couplings.
For example, for $x \sim 0.3$ \cite{Casalbuoni:2003cs}, one finds
$\Delta_A \sim 0.2\, \Delta^B_A$ for $y= 3/2$, while for $x \sim 1$
then $\Delta_A \sim 0.5 \,\Delta^B_A$.

All the gaps feel the presence of the external magnetic field. As
expected, the effect of the magnetic field in $\Delta^{B}_{A}$ is
to increase the density of states, which enters in the argument of
the exponential as typical of a BCS solution. The density of
states appearing in (\ref{gapBA}) is just the sum of those of
neutral and charged particles participating in the given gap
equation (for each Landau level, the density of states around the
Fermi surface for a charged quark is $\widetilde{e}\widetilde{B}/2
\pi^2$).

All the $\widetilde{Q}$-charged quarks have common gap
$\Delta^{B}_{A}$. Hence, the densities of the charged quarks are
all equal. As two of these quarks have positive $\widetilde{Q}$
charge, while the other two have it negative, the $\widetilde{Q}$
neutrality of the medium is guaranteed without having to introduce
any electron density.

Our zero-temperature results imply that a propagating rotated
photon with energy less than the lightest charged quark mode
cannot scatter, since all the $\widetilde{Q}$-charged quarks
acquire a gap and all the Nambu-Goldstone bosons are neutral.
The anisotropy present in the background of an external magnetic field and the
 existence of charged Goldstone bosons in CFL but not in MCFL
indicates a rather different low energy physics, including transport
properties. In particular,  the MCFL superconductor is transparent and
behaves at $T=0$ as
an anisotropic
dielectric, as opposed to the isotropic dielectric behavior of the
CFL phase \cite{Litim:2001mv,Manuel:2001mx}. However, similar to the
CFL, the medium will become optically opaque as soon as leptons
are thermally excited \cite{Shov-Ellis03}.

In previous analysis (see first paper in \cite{magnetic-orig}) the
critical magnetic field at which the CFL pairing is destroyed was
estimated to be $\sim 10^{20}$G. This estimate was based in a
field-independent CFL pairing energy
$\mu^{2}(\Delta^{CFL}_{A})^{2}$. Considering that in the MCFL
phase $\Delta^{B}_{A}$ increases with the field, it is natural to
expect that the critical field will be even larger than
$10^{20}$G. Since such extremely strong fields will surpass all
the energy scales of the system, the quark infrared dynamics will
become predominant, and the  phenomenon of magnetic catalysis of
chiral symmetry breaking \cite{MC} will be activated, producing a
phase with quark-antiquark condensates but no quark-quark
condensate. These two phases will have to be connected by a phase
transition, as they have different number of Goldstone bosons due to the
breaking of baryon symmetry, which only occurs in the
superconducting phase.

Let us stress that in this work we have not considered the
implications of finite quark masses. This, together with a careful
study of the effects of the magnetic field in the low energy
physics, in transport properties or in neutrino dynamics will be
the subject of future investigations.

In conclusion, we have found that a magnetic field leads to the
formation of a new color-flavor locking phase, characterized by a
smaller vector symmetry than the CFL phase. The essential role of
the penetrating magnetic field is to modify the density of states
of charged quarks on the Fermi surface. To better understand the
relevance of this new phase in astrophysics we need to explore the
region of moderately strong magnetic fields
$\widetilde{e}\widetilde{B}< \mu^2/2$, which requires to carry out
a numerical study of the gap equations including the effect of
higher Landau levels. Because the total density of states around
the Fermi surface for charged particles does not vary
monotonically with the number of Landau levels, we still expect to
find a meaningful splitting of the gaps at these fields and
therefore a qualitative separation between the CFL and MCFL
phases.

{\bf Acknowledgments:} We are grateful to M. Alford, V. P.
Gusynin, G. Martinez-Pinedo, K.~ Rajagopal and I.~A.~ Shovkovy for useful
discussions. We thank IEEC for hospitality during completion of
this work. The work of E.J.F. and V.I. was supported in part by
NSF grant PHY-0070986, and C.M. was supported by MEC under grant
FPA2004-00996.


\begin{thebibliography}{}
\bibitem{footnote}We will use the conventions $Q = -e\lambda_8/\sqrt{3}$,
where $\lambda_8$ is the eigth Gell-Mann matrix. We thus use an
unusual ordering in flavor space $(s,d,u)$.
\end{thebibliography}
\end{document}